\documentclass[12pt,amsfonts]{article}    
\usepackage{amsfonts}
\usepackage{CJK}

\def\one{1\hskip-.37em 1}

\def\vp{\varphi}

\def\th{\theta}

\def\si{\sigma}

\def\ra{\rightarrow}

\def\o{\overline}

\def\b{\begin{eqnarray*}}  %takes no eqn numbers
\def\e{\end{eqnarray*}}    %takes no eqn numbers
\def\bn{\begin{eqnarray}}  %takes eqn numbers
\def\en{\end{eqnarray}}   %takes eqn numbers

\def\<{\langle}
\def\>{\rangle}

\def\no{\nonumber}

\def\{{\lbrace}

\def\}{\rbrace}
\begin{document}
\begin{CJK*}{GB}{}

\title{The Favored Classical Variables to \\Promote to Quantum Operators}          % is simple?                    %Choosing an $\hbar$-Counterterm \\ for %the Hamiltonian operator\\in               %Gravity} %{Nonrenormalizable Quantum Field Theories\\With %Relevance for  %Gravity} %Quantum Gravity Made Simple %Easy%
\author{John R. Klauder\footnote{klauder@phys.ufl.edu} \\
Department of Physics and Department of Mathematics \\
University of Florida,   %P.O. Box 118440\\
Gainesville, FL 32611-8440}
\date{ }
%let\frak\cal
\bibliographystyle{unsrt}

\maketitle 
\end{CJK*}
\begin{abstract}
Classical phase-space variables are normally chosen to promote to quantum operators in order to
quantize a given classical system. While classical variables can exploit coordinate transformations
to address the same problem, only one set of quantum operators to address the same problem can give the correct analysis. Such a choice leads to the need to find the favored classical variables in order to achieve a valid quantization. This article addresses the task of how such favored variables are found that can be used to properly solve a given quantum system. Examples, such as non-renormalizable 
scalar fields and gravity, have profited by initially changing which classical variables to promote to quantum operators.
\end{abstract}

\section{A Brief Story of Three Valid Quantization Procedures}
\subsection{Canonical quantization}
Conventional phase-space variables, such as $p$ and $q$, where $-\infty <p,q<\infty$, with Poisson
brackets $\{q,p\}=1$, are natural candidates to promote to basic quantum operators in the procedures that canonical quantization employs. However, traditional coordinate transformations, such as $p\ra 
\o{p}$ and $q\ra\o{q}$, with $-\infty<\o{p},\o{q}<\infty$ and Poisson brackets $\{\o{q},\o{p}\}=1$
are also qualified, in principle, as natural candidates to promote to basic quantum operators. This article resolves the problem of deciding which pair of classical variables to promote to basic quantum operators in order to achieve a valid quantization. 

Dirac gave us the clue about which classical variables to choose for canonical quantization \cite{dirac}, although he did not prove his proposal. Dirac proposed that the proper choice of variables were those that were `Cartesian coordinates'.
This is not obvious because phase space does not have a metric to exhibit Cartesian coordinates.
To fulfill Dirac's rule a metric space that admits Cartesian coordinates must be added to these
particular coordinates. This request requires that the metric of that two-dimensional space must
be given by an expression such as $d\si(p,q)^2= \omega^{-1} \,dp^2+\omega\,dq^2$ for some constant $\omega$.
This space admits a simple shift of coordinates, e.g., $p\ra p+a$ and $q\ra q+b$, where $a$ and $b$ are constants. Moreover, and this is a property that we wish to feature, the shift of these two variables has brought one to a different point on the flat surface that is identical in its surroundings as it was before the shift occurred. Clearly the identity in surroundings for a two-dimensional flat surface
is a property that distinguishes it from almost all other surfaces that are not completely flat. 
Also the Poisson brackets $\{q+b, p+a\}=\{q, p\}=1$ are automatically fulfilled after the shift in location.
If we denote the promotion to quantum operators, e.g., $p\ra P$ and $q\ra Q$, it also follows that 
$p+a\ra P+a\one$ and $q+b\ra Q+b\one$, which leads to $[Q+b\one,P+a\one]=[Q,P]=i\hbar\one$. In brief,
a flat surface and the choice of Cartesian coordinates, with or without $a$ and $b$ shifts, leads to acceptable classical variables to promote to become the basic quantum operators. 

The only problem deals with how that flat surface, which is linked to both the classical and quantum realms, comes about. As a potential link
to other forms of quantization (the spin and affine versions), we call our flat, 
two-dimensional surface a `constant zero curvature'.

\subsubsection{The role of canonical coherent states}
We choose to name the favored classical variables that are promoted to quantum operators as $p$
and $q$, and the valid quantization of these variables named as $P$ and $Q$. We confirm this choice by
first introducing canonical coherent states given by
   \bn |p,q\>\equiv e^{-iqP/\hbar}\,e^{ipQ/\hbar}\,|\omega\>\;,\en
   where $(Q+iP/\omega)\,|\omega\>=0$. It follows that $\<p,q|P|p,q\>=\<\omega|(P+p\one)|\omega\>=p$ and $\<p,q|Q|p,q\>=\<\omega|(Q+q\one)|\omega\>=q$, which is a clear connection between the quantum and classical {}
   basic variables. We finalize this connection with  a Fubini-Study metric \cite{77} that 
   involves a tiny ray distance (minimized over non-dynamical phases) between two infinitely close canonical coherent states given by
   \bn &&d\sigma(p,q)^2\equiv2\hbar[\,|\!|\,d|p,q\>|\!|^2-|\<p,q|\,d|p,q\>|^2]\no \\
   &&\hskip4.1em=\omega^{-1}\, dp^2+\omega\,dq^2\;.\en
   
Observe that this process has given us Cartesian coordinates. These variables clearly are invariant even if we choose $p+a$ and $q+b$. These favored coordinates are Cartesian coordinates and they are promoted to valid basic quantum operators, as Dirac had predicted. 

\subsection{Spin quantization}
The surface of an ideal three-dimensional ball is two-dimensional and spherical with a constant radius; we can say that it has a `constant positive curvature'. Again, like a flat space,
the properties at any point on the spherical surface are exactly like those at any other point on the surface. This is the space on which the spin variables appear. There are three spin operators, $S_1$, $S_2$, and $S_3$, which belong to the groups $SO(3)$ or $SU(2)$. These operators obey certain rules, such as
$[S_1,S_2]=i\hbar\,S_3$, and natural permutations, as well as $S_1^2+S_2^2+S_3^2=\hbar^2 s(s+1)\one_{2s+1}$,
where $2s+1$ is the dimension of the vectors, and the spin $s$ values are $(1,2,3,\ldots)/2$. Some 
basic vectors are $|s,m\>$, where $S_3|s,m\>=m|s,m\>$ and $-s\leq m\leq s$,   $(S_1+iS_2)|s,m\>=|s,m+1\>$, as well as $(S_1+iS_2)|s,s\>=0$.

\subsubsection{The role of spin coherent states}
The spin coherent states are given by
   \bn |\th,\vp\>\equiv e^{-i\vp S_3/\hbar}\,e^{-i\th S_2/\hbar}\,|s,s\>\;, \en
where $-\pi<\vp\leq\pi$, and $0\leq\th\leq\pi$. We also introduce $q=(s\hbar)^{1/2}\,\vp$
and $p=(s\hbar)^{1/2}\,\cos(\th)$. It follows that
  \bn &&d\si(\th,\vp)^2\equiv 2\hbar\,[\,|
  \!|\,d|\th,\vp\>|\!|^2-|\<\th,\vp|\,d|\th,\vp\>|^2\,] \no \\
  &&\hskip4.3em =(s\hbar)[d\th^2+\sin(\th)^2\,d\vp^2\,
  ]\;,\label{ff} \en
  or we can say that
  \bn d\si(p,q)^2\equiv 2\hbar\,[\,|\!|\,d|p,q\>|\!|^2-|\<p,q|\,d|p,q\>|^2\,] \no \\
  &&\hskip-16em =(1-p^2/s\hbar)^{-1}
  dp^2+(1-p^2/s\hbar)\,dq^2\;. \label{ss} \en
  Equation (\ref{ff})makes it clear that we are dealing with a spherical surface with a radius of $(s\hbar)^{1/2}$. Equation (\ref{ss}) makes it clear that if $s\ra\infty$, in which case both  $p$ and $q$ span the real line, we will have recovered properties of canonical quantization.
  
  So far we have obtained surfaces with a constant zero curvature and a constant positive curvature. Could there be more? Could there be surfaces with a `constant negative curvature'?
  
  \subsection{Affine quantization}
  One of the simplest problems to quantize is an harmonic oscillator for which
   $-\infty<p,q<\infty$,
  but it is not so simple if $0<q<\infty$. To solve the second version requires a new method of quantization called affine quantization, which, as we will discover, involves a constant negative curvature. To introduce this procedure let us focus on the classical term $p\, dq$ which is part of a classical action functional. Instead of these variable's range being $-\infty<p,q<\infty$, let us assume 
  $q$ is limited to $0<q<\infty$, and we want to change variables. As a first step, let us consider 
  $ p\, dq=pq\, dq/q=pq\,d\ln(q)$. While $q$ must be positive, $\ln(q)$ covers the whole real line.
  Although that $p\ra pq$ and $q\ra\ln(q)$ both cover the whole real line, we instead just choose $pq$ and $q$ as our new variables. A potential quantization of this pair of variables could involve $q\ra Q$, with $0<Q<\infty$, and $pq\ra(PQ+QP)/2\equiv D$. Note, if $0<Q<\infty$, then $P$ cannot be self adjoint; however, thanks to $Q$, $D$ can be self adjoint, which is a very important advantage. The 
  two basic operators for affine quantization then are $D$ and $Q$, for 
  which $[Q,D]=i\hbar\,Q$.\footnote{Besides $0<q,Q<\infty$, one may also consider $-\infty<q,Q<0$, as
  well as $-\infty<q\neq0,Q\neq0<\infty$.}
 
 \subsubsection{The role of affine coherent states}
 The affine coherent states, where both $q$ and $Q$ have been chosen dimensionless for simplicity, are given by
      \bn |p;q\>\equiv e^{ipQ/\hbar}\,e^{-i\ln(q)\,D/\hbar}\,|b\>\;,  \en
   where $[(Q-1)+iD/b\hbar]\,|b\>=0$. For these variables we find that $\<p;q|Q|p;q\>=
   \<b| qQ|b\>=q$ and $\<p;q|D|p;q\>=\<b| D+pqQ|b\>=pq$. It follows that
      \bn &&d\si(p,q)^2\equiv 2\hbar\,[\,|\!|\,d|p;q\>|\!|^2-|\<p;q|\,d|p;q\>|^2\,] \no   \\
      &&\hskip4.1em =(b\hbar)^{-1}\,q^2\,dp^2 +(b\hbar)\,q^{-2}\,dq^2 \;. \en
      This expression for the Fubini-Study metric is that of a constant negative curvature, an amount of $-2/(b\hbar)$, which is also geodetically complete \cite{p}. Surfaces that are constant negative curvature are only visible in our three-dimensional space at the `center point' only, namely at
      $q=1$, where it appears that in one direction the surface goes down and 90 degrees away the surface goes up. In reality that is what happens at all points for a constant negative curvature,
      but we can not see that effect. Let us give a $q\ra q+b'$ test, with $b'>0$; we use $b'$ so as 
      to make sure this is different from the $b$ that labels the fiducial vector. In that case, we now have
      $(b\hbar)^{-1}(q+b')^2 dp^2+(b\hbar)(q+b')^{-2}dq^2$. This version now implies that  $-b'<q<\infty$.
      Indeed, we can let $b'\ra\infty$ and at the same time, let the factor $b$ in the fiducial vector
      become large ($b\propto b'^2$) in which case we can arrange that the negative curvature $-2/b\hbar\ra0$,
      in such a way that the Fubini-Study metric effectively passes to $B^{-1}dp^2+B\,dq^2$, namely, the constant zero curvature case, with a non-dynamical positive $B$. Basically, we have effectively changed the affine operators $Q$ and $D$ to canonical operators, by 
      $Q\ra Q+b'\one$ and $D\ra D+b'P$ so that, as $b'\ra\infty$, 
       $[Q+b'\one, D+b'P]/b'=i\hbar(Q+b'\one)/b'\ra
      [Q,P]=i\hbar\one$.
      
      \section{Summary}
      Our quantization of classical variables has now been completed. It was shown that classical variables that represent the coordinates of constant positive, zero, and negative curvatures,
      complete the natural forms of surface and these three divisions include a different variety of
      classical systems that can be quantized. Affine quantization, as a special procedure to quantize systems, has not yet become universally well known and exploited; it deserves more attention. 
      Besides the harmonic oscillator with $0<q<\infty$ (see \cite{t}), there are other 
      problems for which affine quantization can work well. Quite
      recently, a surprisingly transparent version of affine quantization has been used for non-renormalizable covariant scalar fields and for Einstein's gravity \cite{k}.
      
      Others are encouraged to see what affine quantization can do for their own 
      quantization problems.

\end{document}